\documentclass[10pt,letterpaper,conference]{IEEEtran}
\usepackage{geometry}                
\usepackage{fullpage}
\usepackage{graphicx}
\usepackage{amssymb}
\usepackage{amsmath}%
\usepackage{amsfonts}%
\usepackage{amssymb}%
\usepackage{amsthm}
\usepackage{breqn}
\usepackage{graphicx}
\usepackage{colonequals}
\usepackage{psfrag}
\usepackage{subfigure}
\usepackage{mathrsfs}
\usepackage{cite}

\newtheorem{prop}{Proposition}

\theoremstyle{definition}

\newtheorem{defn}{Definition}
\newtheorem{scheme}{Scheme}
\newtheorem{remark}{Remark}

\newcommand{\etal}{\textit{et al.}~}
\newcommand{\Fq}{\mathbb{F}_{q}}
\newcommand{\PR}{\mbox{Pr}}

\newcommand{\calX}{\mathcal{X}}
\newcommand{\calJ}{\mathcal{J}}
\newcommand{\calC}{\mathcal{C}}
\newcommand{\calS}{\mathcal{S}}

\newcommand{\Xn}{\mathcal{X}^n}
\newcommand{\cwd}{\{1,\dots,2^{nR} \}}
\newcommand{\floor}[1]{\lfloor #1 \rfloor}
\newcommand{\ceil}[1]{\lceil #1 \rceil}

\newcommand{\bc}{\mathbf{c}}
\newcommand{\bH}{\mathbf{H}}

\newcommand{\bD}{\mathbf{D}}
\newcommand{\bx}{\mathbf{x}}
\newcommand{\rank}{\mbox{rank}}
\newcommand{\bz}{\mathbf{z}}
\newcommand{\bX}{\mathbf{X}}
\newcommand{\bY}{\mathbf{Y}}

\newcommand{\Enc}{\mathsf{Enc}}
\newcommand{\Dec}{\mathsf{Dec}}
\IEEEoverridecommandlockouts

\title{Lists that are smaller than their parts: \\ A coding approach to tunable
secrecy  \thanks{This work is sponsored by the Department of Defense under Air
Force Contract FA8721-05-C-0002. Opinions, interpretations, recommendations, and
conclusions are those of the authors and are not necessarily endorsed by the
United States Government. Specifically, this work was supported by Information
Systems of ASD(R\&E).\newline $^{\ddagger}$ Currently
with Auroral LLC. \newline $^{**}$ Supported by the Irish Higher Educational
Authority (HEA) PRTLI
Network Mathematics Grant.} }
\author{\IEEEauthorblockN{Fl\'avio du Pin Calmon$^\dagger$, Muriel
M\'edard$^\dagger$, Linda M. Zeger$^\ddagger$,\\ Jo\~ao Barros$^*$,  Mark M. Christiansen$^{**}$, Ken R. Duffy$^{**}$}
\IEEEauthorblockA{$^\dagger$Massachusetts Institute of Technology, Cambridge, MA, \{flavio, medard\}@mit.edu\\
$^\ddagger$MIT Lincoln Laboratory, Lexington, MA, zeger@auroral.biz\\
$^*$ Instituto de Telecommunica\c{c}\~oes, FEUP, Porto, Portugal, jbarros@fe.up.pt\\
$^{**}$ Hamilton Institute, National University of Ireland, Maynooth, Ireland, \{mark.christiansen, ken.duffy \}@nuim.ie 
}

}

\begin{document}

\maketitle
\vspace{-2in}

\begin{abstract}
  We present a new information-theoretic definition and  associated results,
  based on list decoding in a source coding setting. We  begin by presenting
  list-source codes, which naturally map a key length  (entropy) to list size.
  We then show that such codes can be analyzed in the  context of a novel
  information-theoretic metric, $\epsilon$-symbol secrecy,  that encompasses
  both the one-time pad and traditional rate-based asymptotic  metrics, but,
  like most cryptographic constructs, can be applied in  non-asymptotic settings.
  We derive fundamental bounds for $\epsilon$-symbol  secrecy and demonstrate
  how these bounds can be achieved with MDS codes when  the source is uniformly
  distributed. We discuss applications and  implementation issues  of our codes. 
\end{abstract}

\section{Introduction }
Classic information-theoretic approaches to secrecy are  concerned with
unconditionally secure systems, i.e.  schemes that manage to hide all the bits
of a message from an adversary with unbounded computational resources.  It is
well known that, for a noiseless setting, unconditional (i.e. perfect) secrecy
can only be attained when both communicating parties share a random key with
entropy at least as large as the message itself
\cite{shannon_communication_1949}.  In other cases, perfect secrecy can
sometimes be achieved by exploiting particular characteristics of the considered
model, such as when the legitimate communicating party has a less noisy channel
than the eavesdropper (wiretap channel) \cite{liang_information_2009}. 


Alternatively, computationally secure cryptosystems have thrived both from a
theoretical and a practical perspective.  Such systems are based on yet unproven
hardness assumptions, but nevertheless have led to cryptographic schemes that
are widely adopted (for an overview, see \cite{katz_introduction_2007}).
Currently, computationally secure schemes are used millions of times per day, in
applications that range from online banking transactions to digital rights
management. However, with the ever increasing amount of data streaming over the
Internet and the need to provide secure connections to mobile low powered
devices, there is still a constant demand for new and efficient security
solutions.


There has been a long exploration of the connection between coding and
cryptography  \cite{blahut_communications_1994}, and our work is inscribed in
this school of thought.  From a theoretical perspective, we aim to present a new
framework that allows the application of information theoretic-tools to analyze
a broader set of secrecy schemes that go beyond the one-time pad and the wiretap
model with its variations. Towards this goal, we  define a new metric for
analyzing  security, namely $\epsilon$-\textit{symbol secrecy}, which
quantifies the uncertainty of specific source symbols given an encrypted source
sequence.  This metric subsumes traditional rate-based information-theoretic
measures of secrecy which, unlike usual cryptographic approaches, are generally
asymptotic. However, our definition is not asymptotic and, indeed, we provide a
construction that achieves fundamental symbol secrecy bounds, based on MDS
codes, for finite-length sequence.

In order  to  construct schemes that achieve  symbol secrecy performance bounds,
we present the definition of \textit{list-source codes}, which are  codes that
compress a source sequence \textit{below} its entropy rate. Consequently, a
list-source code is decoded to a list  of possible source sequences instead of a
unique source sequence.  Fundamental bounds for list-source codes are derived,
and explicit constructions that achieve such bounds are presented using tools
from algebraic coding theory.  

We show how list-source codes can be used as an important tool for hiding
information with key sizes that are only a fraction of the entropy of the
message. Using list-source codes, it becomes possible to argue that the best an
adversary can do is to reduce the set of possible messages to an exponentially
sized list with certain properties, where the size of the list depends on the
length of the key. Since the list has an exponential size, it cannot be resolved
in polynomial time, offering a certain level of computational security. We will
show how this property can be used to develop hybrid encryption schemes, where
only part of the message needs to be securely encrypted.

 Our main practical application of interest is  secure content caching and
 distribution. We propose a hybrid encryption scheme based on list-source codes,
 where a large fraction of the message can be encoded and distributed using a
 key-independent list-source code. The information necessary to resolve the
 decoding list, which can be much smaller than the whole message, is then
 encrypted using a secure  method. This scheme allows a significant amount of
 content to be distributed and cached  \textit{before} dealing with key
 generation, distribution and management issues.  

\subsection{Related work}

Tools from algebraic coding theory have been widely used for constructing
secrecy schemes  \cite{blahut_communications_1994}. In addition, the notion of
providing security by exploiting the fact that the adversary has incomplete
access to information is also central to several secure network coding schemes
and wiretap models. Ozarow and Wyner \cite{ozarow_wire-tap_1985} introduced the
wiretap channel II, where an adversary can observe a  set $k$  of his choice out
of $n$ transmitted symbols, and proved that there exists a code  that achieves
perfect secrecy. A generalized version of this model was investigated by Cai and
Yeung in \cite{cai_secure_2002}, where they introduce  the related problem of
designing an information-theoretically secure linear network code when an
adversary can observe a certain number of edges in the network. Their results
were later extended in
\cite{feldman_capacity_2004,mills_secure_2008,el_rouayheb_secure_2012,silva_universal_2011}.
A more practical approach was presented by Lima \etal in
\cite{lima_random_2007}. For a survey on the theory of secure network coding, we
refer the reader to \cite{cai_theory_2011}.

The setting considered in this paper is related to the wiretap channel II in
that a fraction of the source symbols is hidden from a possible adversary.
Oliveira \etal investigated in \cite{oliveira_trusted_2010} a related setting in
the context of data storage over untrusted networks that do not collude,
introducing a solution based on Vandermonde matrices. The MDS coding scheme
introduced in this paper is similar to \cite{oliveira_trusted_2010}, albeit the
framework developed here  is more general.

List decoding techniques for channel coding were first introduced by Elias
\cite{elias_list_1957} and Wozencraft \cite{wozencraft1958}, with subsequent
work by Shannon \etal \cite{shannon_lower_1967,shannon_lower_1967-1}  and Forney
\cite{forney_exponential_1968}. Recently, new algorithmic results for list
decoding of channel codes were discovered by Gurusuwami and Sudan
\cite{guruswami_list_2001}. We refer the reader to \cite{guruswami_list_2009}
for an excellent survey of  list decoding results. List decoding has been
considered in the context of source coding in \cite{ali_source_2010}. The
approach is related to the one presented here, since we may view a secret key as
side information, but \cite{ali_source_2010} do not consider source coding and
list decoding together for the purposes of security.

\subsection{Communication and threat model}
\label{sec:model}

A transmitter (Alice)  sends to a legitimate receiver (Bob) a sequence of length
$n$ produced by a discrete  source $X$ with output alphabet $\mathcal{X}$ and
probability distribution $p_{X}(\cdot)$. Both Alice and  Bob have access to a
shared secret key $K$ drawn uniformly and at random from a discrete alphabet
$\mathcal{K}$, such that $H(K)<H(X^n)$, and encryption/decryption functions
$\Enc:\calX^n\times \mathcal{K} \rightarrow \mathcal{C}$ and
$\Dec:\mathcal{C}\times \mathcal{K}\rightarrow \calX^n$, where $\mathcal{C}$ is
the set of possible encrypted messages. In addition, Alice communicates with Bob
over a noiseless channel. Alice observes the source sequence $X^n$, and
transmits an encrypted message $C= \Enc(X^n, K)$. Bob then recovers $X^n$ by
decrypting the message using the key, recovering $\hat{X}^n=\Dec(C,K)$. The
communication is successful if $\hat{X}^n=X^n$.

We assume a passive but computationally unbounded eavesdropper (Eve) that has
access to all transmitted messages from Alice to Bob and knows the  functions
$\Enc(\cdot)$ and $\Dec(\cdot)$, but does not know the secret key $K$. Eve's
goal is to gain as much knowledge as possible about the original source
sequence.  This is the traditional framework used in cryptography, and no
degraded assumption is made beyond the shared secret key.

In the remainder of this paper we investigate two main aspects of this model,
described below.

\subsubsection{Encryption with key entropy smaller than the message entropy} 
We initially analyze how to perform encryption when the key is smaller than the
message. Towards this goal, we present the definition of list-source codes
(LSCs), together with fundamental bounds, in section  \ref{sec:LSC}.
Furthermore, practical code constructions of LSCs are introduced in section
\ref{sec:code}. We present list-source codes as codes that compress the source
sequence \textit{below} its entropy rate, and in section \ref{sec:code} describe
how LSCs can be  used in the considered model.

\subsubsection{Security analysis and new security metrics for i.i.d. sources} We
analyze the security of  schemes based on LSCs in section \ref{sec:security}. In
addition, we introduce a new information-theoretic metric that can be used in
scenarios where perfect secrecy cannot be achieved, namely \textit{absolute} and
$\epsilon$\textit{-symbol secrecy}.

In section \ref{sec:general} we discuss the extension of LSCs to Markovian
source models, and in section \ref{sec:practical} we present applications and
practical considerations of the proposed secrecy scheme. Finally, section
\ref{sec:conc} presents our concluding remarks.

\section{List decoding and source coding: Fundamental Limits}
\label{sec:LSC}

In this section we present the definition of list-source codes and derive
fundamental bounds. Consider a discrete memoryless source $X$ with output
alphabet $\mathcal{X}$
and probability distribution $p_{X}(\cdot)$. 
\begin{defn}
  A $(2^{nR},|\mathcal{X}|^{nL},n)$-list-source code for a discrete memoryless  source $X$ consists
of an encoding function $f_n:\mathcal{X}^n\rightarrow\cwd$ and a list-decoding
function $g_n:\cwd\rightarrow \mathcal{P}(\Xn)\backslash \varnothing$, where
$\mathcal{P}(\Xn)$ is the power set of $\Xn$ and 
$|g(w)|= |\mathcal{X}|^{nL}~\forall w\in\cwd$.
\end{defn}

 Note that $0\leq L \leq 1$. From an operational point of view, $L$ is a
 parameter that determines the size of the decoded list. For example, $L=0$
 corresponds to traditional lossless compression, i.e., each source sequence is
 decoded to a unique sequence. Furthermore, $L=1$ represents the trivial case
 when the decoded list corresponds to $\Xn$. 
 
 For a list-source code, an error is declared
when a string generated by a source is not contained in the corresponding
decoded list. The average error probability is given by
 \begin{equation}       
  e_{L}(f_n,g_n)=\PR(X^n\notin g_n(f_n(X^n))).
\end{equation}
\begin{defn} 
  For a given discrete memoryless source $X$, the rate list size pair $(R,L)$ is said to be
  \textit{achievable} if for every $\delta>0$, $0<\epsilon<1$ and sufficiently
  large $n$ there exists a sequence of $(2^{nR_n},|\mathcal{X}|^{nL_n},n)$-list-source codes
  $(f_n,g_n)$ such that $R_n< R+\delta$, $|L_n- L|< \delta$ and $e_{L_n}(f_n,g_n)\leq \epsilon$. The
  \textit{rate list region} is the closure
  of all rate list pairs $(R,L)$.
\end{defn}

\begin{defn}
  The \textit{rate list function} $R(L)$ is the infimum of all rates $R$
such that $(R,L)$ is in the rate list region for a given normalized list size
$0\leq L \leq 1$.
\end{defn}

\begin{prop}
For any discrete memoryless source X, the rate list function is bounded below by
\begin{equation} 
\label{eq:ratelist}
  R(L)\geq H(X)-L\log|\mathcal{X}|~.
\end{equation}
\end{prop}

\begin{figure}[!tb]
  \begin{center}
    \psfrag{1}[c][c]{$R$}
    \psfrag{2}[b][l][1][90]{$L$}
    \psfrag{3}[r][r]{\small$\displaystyle  \frac{H(X)}{\log|\mathcal{X}|}$}
    \psfrag{4}[c][c]{\small$0$}
    \psfrag{5}[c][c]{\small$0$}
    \psfrag{6}[c][c]{\small$H(X)$}
    \psfrag{7}[l][t]{Achievable}
    \psfrag{8}[c][c]{1}
    \includegraphics[scale=0.42]{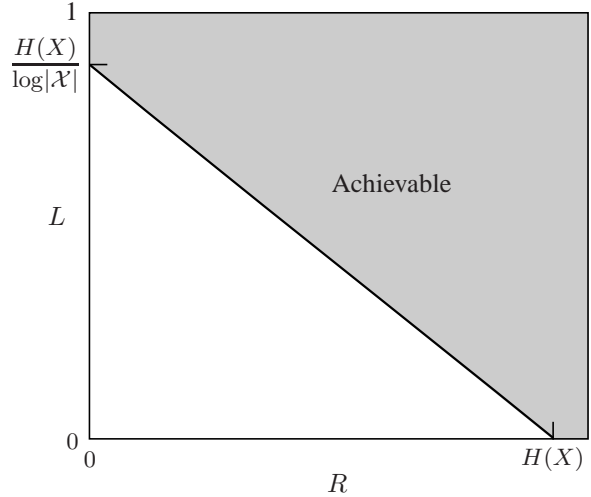}
  \end{center}
  \caption{Rate list region for normalized list size $L$ and code rate $R$.}
  \label{fig:conj}
\end{figure}

\begin{proof}
Let $\delta>0$ be given and $(f_n,g_n)$ be a sequence of codes with
(normalized) list size $L_n$ such
that  $L_n\rightarrow L$  and for any $0<\epsilon<1$ and $n$ sufficiently large $0\leq e_L(f_n,g_n)\leq \epsilon$. Then 
\begin{align} 
  \PR \left[X^n\in \displaystyle \bigcup_{w\in \mathcal{W}^n} g_n(w) \right]
  &\geq \PR[X^n\in g_n(f_n(X^n))]\\
  &\geq 1-\epsilon
\end{align}
where $\mathcal{W}^n=\{1,\dots,2^{nR_n} \}$ and
$R_n$ is the rate of the code $(f_n,g_n)$. Using \cite[Lemma 2.14]{csiszar_information_2011}:
\begin{align}
  \frac{1}{n} \log\left( \displaystyle \sum_{w\in \mathcal{W}^n} |g_n(w)|\right)&=\frac{1}{n}
  \log\left(  2^{nR_n}|\mathcal{X}|^{nL_n} \right) \nonumber\\
  &= R_n+L_n\log |\mathcal{X}|\nonumber \\
  &\geq  \frac{1}{n} \log  \left| \bigcup_{w\in
  \mathcal{W}^n} g_n(w)  \right| \nonumber \\ 
  &\geq H(X)-\delta
\end{align}
if $n\geq n_0(\delta,\epsilon,|\mathcal{X}|)$. Since this holds for any
$\delta>0$, it follows that $R(L)\geq H(X)-L\log |\mathcal{X}|$ for all $n$ sufficiently large.

\end{proof}
\begin{remark}
Achievability of the bound \eqref{eq:ratelist}  will be shown through an explicit design using linear codes in the next section, so the inequality can be proved to be an equality.
\end{remark}

\section{Code design}
\label{sec:code}

\subsection{Trivial approach}
Assume that the source $X$ is uniformly distributed in $\mathbb{F}_q$, i.e.,
$\PR(X=x)=1/q~\forall x \in \mathbb{F}_q$. In this case $R(L)=(1-L)\log q$. A trivial
scheme for achieving the list-source boundary is the following. Consider a
source sequence   $X^n=(X^p,X^s)$, where $X^p$ denotes the first
$p=n-\floor{Ln}$ symbols of $X^n$ and $X^s$ denotes the last $s=\floor{Ln}$
symbols. Encoding is done by discarding $X^s$, and mapping the prefix  $X^p$ to
a binary codeword $Y^{nR}$ of length $nR=\ceil{n-\floor{Ln}\log q}$ bits.

For decoding, the codeword $Y^{nR}$ is mapped to $X^p$, and the scheme outputs a
list of size $q^s$ composed by $X^p$ concatenated with all possible combinations
of  suffixes of length $s$. Clearly, for $n$ sufficiently large, $R\approx
(1-L)\log q$, and we achieve the optimal list-source size tradeoff. 

The previous scheme is completely inadequate for security purposes. An adversary that
observes the binary codeword $Y^{nR}$ can uniquely identify the first $p$
symbols of the source message, and the uncertainty is  concentrated over the
last $s$ symbols. Ideally, assuming that all source symbols are of equal
importance, we should  spread the uncertainty over all symbols of the message.
More precisely, given the encoding $f(X^n)$, a ``good''  security scheme would
provide $I(X_i;f(X^n)) \leq \epsilon\ll \log q$ for $1\leq
i \leq n$. Of course, we can naturally extend this notion for groups of symbols
or functions over input symbols\footnote{This idea is tightly related to the
concept of hard core predicates and semantic security in cryptography. }. This
idea will be captured in the definition of \textit{symbol secrecy}, introduced
in section \ref{sec:security}.

\subsection{A construction based on linear codes}

 Let $X$ be an i.i.d. source with $X\in \calX$ with entropy $H(X)$, and
 $\calS_n$ a   source code with  encoder $s_n:\calX^n \rightarrow \Fq^{m_n}$
 and decoder  $r_n:\Fq^{m_n}\rightarrow \calX^n$. Furthermore, let
 $\mathcal{C}$ be a $(m_n,k_n,d)$ linear code over $\Fq$ with an
 $(m_n-k_n)\times m_n$ parity check matrix $\mathbf{H}_n$ (i.e. $\bc \in
 \mathcal{C} \Leftrightarrow \bH_n \bc =0 $).  Consider the following scheme,
 where $k_n=n L_n \log |\calX|/\log q$ for $0\leq L_n \leq 1$ and $L_n \rightarrow L$ as
 $n\rightarrow \infty$. To simplify notation, we assume without loss of
 generality that $k_n$ is an integer.

\begin{scheme}
\label{scheme:lin}
\textit{Encoding}: Let $X^n$ be the  sequence generated by the source.  Compute
the syndrome $S^{m_n-k_n}=\bH_n s_n(X^n)$ and map  each syndrome to a
distinct  sequence of $nR=\ceil{(m_n-k_n)\log q}$ bits, denoted by
$Y^{nR}$.

\textit{Decoding}: Map the binary codeword  $Y^{nR}$ to the corresponding
syndrome $S^{m_n-k_n}$. Output $r_n(x^{m_n})$ for each $x^{m_n}$ in
the coset of $\bH_n$ corresponding to $S^{m_n-k_n}$.
\end{scheme}

\begin{prop}
\label{prop:achieve}
If $\calS_n$ is asymptotically optimal for source $X$, i.e. $
m_n/n\rightarrow H(X)/\log q$, scheme \ref{scheme:lin} achieves the optimal
list-source tradeoff point $R(L)$ for an i.i.d. source, where $R(\cdot)$ is the
rate list function.
\end{prop}
\begin{proof} 
Since the size of each coset corresponding to a syndrome a $S^{m_n-k_n}$ is
exactly $q^{k_n}$, the normalized list size is $L_n=(k_n\log q)/(n\log |\calX|)\rightarrow
L $. Denoting $ m_n/n= H(X)/\log q+\delta_n$, where $\delta_n\rightarrow
0$, it follows that   is $R=\ceil{(m_n-k_n)\log
q}/n=\ceil{(H(X)+\delta_n\log q)n-L_n n \log |\calX|}/n$, which is arbitrarily close
to the rate in \eqref{eq:ratelist} for sufficiently large $n$.
\end{proof}


The source coding scheme used in the proof of Proposition $\ref{prop:achieve}$ can be any asymptotically optimal scheme. Note that if the source $X$ is uniform, and assuming without loss of generality that $L_n=L$ and that $Ln$ is an integer, any message in the coset  of $\mathcal{C}$ determined by $S^{(1-L)n}$ is
equally likely. Hence, $H(X^n|S^{(1-L)n})=q^{L n}$. Scheme \ref{scheme:lin}
provides a systematic way of hiding information, and we can take advantage of
the properties of the underlying linear code to make precise assertions
regarding the ``information leakage'' of the scheme.

With the syndrome in hand, how can we recover the rest of the message? One
possible approach is to find a $k \times n$ matrix $\bD$ that has full rank such
 that the rows of $\bD$ and $\bH$ form a basis of $\Fq^n$. Such a matrix can
be easily found, for example, using the Gram-Schmidt process with the rows of
$\bH$ as a starting point. Then we simply calculate $T^{L n}=\bD X^n$ and
forward $T^{L n}$ to the receiver. The receiver can then invert the system
\begin{equation}
\left( 
\begin{array}{c}
\bH\\
\bD
\end{array}
 \right)
 X^n=\\
 \left(\begin{array}{c}
S^{(1-L )n}\\
T^{L n}
\end{array}\right),
\end{equation}
and recover the original sequence $X^n$. This property allows list-source codes
to be deployed in practice using well known linear code constructions, such as
Reed-Solomon or LDPC.

\begin{remark}
This approach is valid for general linear spaces, and holds for any pair of full
rank matrices $\bH$ and $\bD$ with dimensions $(n-k)\times n$ and $k\times n$,
respectively, such that $\rank ([\bH^T \bD^T]^T)=n$. However, here we adopt the
nomenclature of linear codes since we  make use of known code constructions to
design secrecy schemes in the following sections.
\end{remark}



\subsection{A secure communication scheme based on list-source codes}

In this section we present a general description of a two-phase secure
communication scheme for the model introduced in section \ref{sec:model},
presented in terms of the list-source code constructions derived using linear
codes. Note that this scheme can be easily extended to any list-source code by
using the corresponding encoding/decoding functions instead of multiplication by
parity check matrices. 

We assume that Alice and Bob have access to a encryption/decryption scheme
$(\mathsf{Enc}',\mathsf{Dec}')$ that is used with the shared secret key $K$ and
is sufficiently secure against the adversary. This scheme can be, for example, a
one-time pad. The encryption/decryption procedure is performed as follows, and
will be used as components of  the overall encryption scheme
$(\mathsf{Enc},\mathsf{Dec})$ described below. 

\begin{scheme}
\label{scheme:prac}

\textit{Input}: The source encoded sequence $X^n\in \Fq^n$, parity check matrix $\bH$ of
a linear code in $\Fq^n$, a full-rank $k\times n$  matrix $\bD$ such that
$\rank([\bH^T~\bD^T])=n$, and encryption/decryption functions
$(\mathsf{Enc'},\mathsf{Dec'})$.

\noindent \textbf{Encryption} $(\Enc)$:

\noindent \textit{Phase I (pre-caching)}: Alice generates $S^{n-k} = \bH X^n$ and  sends to Bob.

\noindent \textit{Phase II (send encrypted data)}: Alice generates $E^k=\mathsf{Enc'}(\bD X^n,K)$ and sends to Bob.

\noindent \textbf{Decryption} $(\Dec)$: Bob calculate $\bD X^n = \Dec'(E^k)$ and recover $X^n$ from $S^{n-k}$ and $\bD X^n $.

\end{scheme}

Assuming that  $(\mathsf{Enc'},\mathsf{Dec'})$ is secure, the security of scheme \ref{scheme:prac} reduces to the security of the underlying list-source code (i.e. scheme \ref{scheme:lin}). In practice, the encryption/decryption functions $(\mathsf{Enc'},\mathsf{Dec'})$ may depend on a secret or public/private key, as long as it provide sufficient security for the desired application.  In addition, assuming that the source sequence is uniform and i.i.d. in $ F_q^n$, we can use MDS codes to make strong security guarantees, as described in the next section. In this case, an adversary that observes $S^{n-k}$ cannot infer \textit{any} information about any set of $k$ symbols of the original message. 

 Note that this scheme
 has a  \textit{tunable} level of secrecy: The amount of data sent in phase I and
phase II can be appropriately selected to match the properties of the encryption
scheme available, the size of the key length, and the desired level of secrecy.
Furthermore, when the encryption procedure has a higher computational cost than
the list-source encoding/decoding operations, list-source codes can be used to
reduce the total number of operations required by allowing encryption of a
smaller portion of the message (phase II).

\section{New metrics for security analysis} 
\label{sec:security}

We introduce a new information-theoretic metric for security called
$\epsilon$-symbol secrecy. This metric can be used to characterize the
properties of  security schemes that do not provide absolute secrecy (such as in scheme
\ref{scheme:prac}). Given a source sequence $X^n$ and its
corresponding encryption $Y$, $\epsilon$-symbol secrecy is  the largest fraction
$t/n$ such that  at most $\epsilon$ bits can be inferred from any $t$-symbol
subsequence of $X^n$. We derive a fundamental bound for $\epsilon$-symbol
secrecy, and show that it can be achieved using MDS codes for $\epsilon =0$ and
uniform i.i.d. sources. Before presenting the definition, we make a few comments
on notation and briefly review the threat model.


\subsection{Notation}

 Let $\calC_n$ be a sequence of codes for a discrete memoryless source $X$ with
 probability distribution $p(x)$ that achieves a rate list pair $(R,L)$.
 Furthermore, let $Y^{nR_n}$ be the corresponding codeword $f_n(X^n)$ created by
 $\calC_n$.  Denote by $\mathcal{I}_n(t)$ the set of all subsets of
 $\{1,\dots,n\}$ of size $t$, i.e. $\calJ\in \mathcal{I}_n(t) \Leftrightarrow
 \calJ\subseteq \{1,\dots,n\}$ and $|\calJ| = t$. In addition, we denote by
 $X^{(\calJ)}$ the set of symbols of $X^n$ indexed by the elements in the set
 $\calJ\subseteq \{1,\dots,n\}$.
 
 As discussed in section \ref{sec:model}, we assume a passive but
 computationally  unbounded adversary that only has access to the list-source
 encoded message $f_n(X^n)=Y^{nR_n}$. Based on the observation of $Y^{nR_n}$,
 the adversary will attempt to determine what is the original message. In
 addition, we assume that the source statistics and the list-source code used
 are universally known, i.e. an adversary has access to the distribution
 $p_{X^n}(X^n)$ of the symbol sequences produced by the source and the sequence
 of codes $\calC_n$. We use the standard information-theoretic approach of
 measuring the amount of information that an adversary can gain of a specific
 sequence of source symbols $X^{(\calJ)}$ by observing $Y^{nR_n}$ as  the mutual
 information $I(X^{(\calJ)};Y^{nR_n})$.
 
 \subsection{Symbol Secrecy}

The following definition introduces two security metrics, namely
\textit{absolute symbol secrecy} and $\epsilon$-\textit{symbol secrecy}.

\begin{defn}

We define $\mu_0(\calC_n)$ as the \textit{absolute symbol secrecy} of a code
$\calC_n$ as 
 \begin{equation}
\mu_0(\calC_n)  = \max \left\{ \frac{t}{n} : I(X^{(\calJ)}; Y^{nR_n})=0,~\forall \calJ\in \mathcal{I}_n(t)\right\}.
\end{equation}
The absolute symbol secrecy $\mu_0$ of a sequence of codes $\calC_n$ is:
\begin{equation}
  \mu_0  = \liminf_{n\rightarrow \infty} \mu_0(\calC_n).
\end{equation}
Furthermore, we define the $\epsilon$-\textit{symbol secrecy} $\mu_\epsilon$ of
a code $\calC_n$  as
 \begin{equation}
   \mu_\epsilon(\calC_n )  = \max \left\{\frac{ t}{n} : \frac{1}{t}I(X^{(\calJ)} ; Y^{nR_n})\leq \epsilon~\forall \calJ\in \mathcal{I}_n(t)\right\},
\end{equation}
and the $\epsilon$-symbol secrecy of a sequence of codes $\calC_n$ as 
\begin{equation}
  \mu_\epsilon  = \liminf_{n\rightarrow \infty} \mu_\epsilon(\calC_n),
\end{equation}
where $\epsilon<H(X)$.
\end{defn}

\begin{prop}
Let $\calC_n$ be a sequence of list-source codes that achieves a rate-list pair
$(R,L)$ and an $\epsilon$-symbol secrecy of $\mu_\epsilon$. Then $0\leq \mu_\epsilon
\leq \min \left\{ \frac{L  \log |\calX|}{H(X)-\epsilon},1\right\}$.
\label{prop:epsilon_bound}
\end{prop}
\begin{proof}
We denote $\mu_\epsilon(\calC_n
)=\mu_{\epsilon,n} $. Note that
\begin{align*} 
  I(X^{(\calJ)};Y^{nR_n})&=H(X^{(\calJ)})-H(X^{(\calJ)}|Y^{nR_n})\\
  &= n\mu_{\epsilon,n}H(X)-H(X^{(\calJ)}|Y^{nR_n})\\
  &\leq n\mu_{\epsilon,n} \epsilon.
\end{align*}
Therefore
\begin{align*} 
  \mu_{\epsilon,n}(H(X)-\epsilon)&\leq \frac{1}{n}H(X^{(\calJ)}|Y^{nR_n})\\
  &\leq L_n\log|\mathcal{X}|.
\end{align*}
The result follows by taking $n\rightarrow \infty$.
\end{proof}

The previous result bounds the amount of information an adversary gains about
particular source symbols by observing a list-source encoded
message. In particular, for $\epsilon=0$, we find a meaningful bound on what is
the largest fraction of input symbols that is
\textit{perfectly} hidden. A simple upper-bound for the  maximum average amount of
information that an adversary can gain from a message encoded with any source
code $\calC_n$ with
symbol secrecy $u_{\epsilon,n}$ is given below.

\begin{prop}
  For any code $\calC_n$ 
  for a discrete memoryless source $X$ and any $\epsilon$ such that $0\leq
  \epsilon \leq H(X)$, we
  have
\begin{align} 
  \label{eq:mueps_rel}
  \frac{1}{n}I(X^{n} ; Y^{nR_n}) \leq H(X) - \mu_{\epsilon,n}(H(X)-\epsilon),
\end{align}
where  $\mu_{\epsilon,n}=\mu_\epsilon({\calC_n})$.
\end{prop}
\begin{proof}
  Let  $\mu_{\epsilon,n}=t/n$, $\calJ \in \mathcal{I}_n(t)$  and
  $\bar{\calJ}=\{1,\dots,n\}\backslash \calJ$. Then
\begin{align}
  \frac{1}{n}I(X^{n} ; Y^{nR_n}) &\leq \frac{t}{n}\left(\epsilon+
  \frac{1}{t}I(X^{(\bar{\calJ})} ; Y^{nR_n}|X^{(\calJ)})   \right)\\
  &\leq \mu_{\epsilon,n}\epsilon +\frac{(n-t)}{n}H(X) \\
  &=  H(X) - \mu_{\epsilon,n}(H(X)-\epsilon).
\end{align}
\end{proof}

The next proposition relates the rate-list function with $\epsilon$-symbol
secrecy through the upper bound in proposition \ref{prop:epsilon_bound}.

\begin{prop}
If a sequence of list-source codes $\calC_n$
 achieves a point $(R',L)$ with $\mu_\epsilon=\frac{L  \log
 |\calX|}{H(X)-\epsilon}$ for some $\epsilon$, where $R'=\lim_{n\rightarrow
 \infty}\frac{1}{n}H(Y^{nR_n})$, then $R'=R(L)$.
\end{prop}
\begin{proof}
Assume that $\calC_n$ satisfies the conditions in the proposition and $\delta>0$
is given. Then for $n$ sufficiently large, we have from \eqref{eq:mueps_rel}:
  \begin{align*}  
    \frac{1}{n}H(Y^{nR_n}) &= \frac{1}{n}I(X^{n} ; Y^{nR_n})\\
    &\leq   H(X) - \mu_{\epsilon}(H(X)-\epsilon)+\delta\\
    & = H(X) - L\log|\calX|+\delta .
  \end{align*}
Since this holds for any $\delta$, then $R'\leq H(X) - L\log|\cal X|$. However,
from proposition 1, $R'\geq H(X) - L\log|\cal X|$, and the result follows. 
\end{proof}

\subsection{A scheme based on MDS codes}

We now prove that for a uniform i.i.d. source $X$ in $\Fq$, using scheme
\ref{scheme:lin} with an MDS parity check matrix $\bH$ achieves $\mu_0$. Since
the source is uniform and i.i.d., no source coding is used.

\begin{prop}
If $\bH$ is the parity check matrix of an $(n,k,d)$ MDS and the source $X^n$ is uniform and i.i.d., then Scheme \ref{scheme:lin} achieves the upper bound $\mu_0 = L$, where $L=k/n$.
\end{prop}
\begin{proof}
Let  $\bH$ be the parity check matrix  of a  $(n,k,n-k+1)$ MDS code
$\mathcal{C}$ over $\Fq$, and let $\bx\in \calC$. Fix a set $\calJ \in
\mathcal{I}_{n}(k)$ of $k$ positions of $\bx$, denoted $\bx^{(\calJ) }$. Since
the minimum distance of $\calC$ is $n-k+1$, for any other codeword in $\bz \in
\calC$ we have  $\bz^{(\calJ)}\neq \bx^{(\calJ)}$. Denoting by
$\calC^{(\calJ)}=\{x^{(\calJ)} \in \Fq^k:x\in \calC \}$, then
$|\calC^{(\calJ)}|=|\calC|=q^k.$ Therefore, $\calC^{(\calJ)}$ contains all
possible combinations of $k$ symbols. Since this property also holds for any
coset of $\bH$, the result follows.

\end{proof}


\section{List-source codes for general source models}
\label{sec:general}

The previous results hold for i.i.d. source models. However, for more general
sources the analysis becomes significantly more convoluted, since multiple
list-source encoded messages can reveal information about each other.
Considering that encryption is performed over multiple blocks of source symbols,
the list size will not necessarily grow if these block are correlated. 


In general, given an output $\bX=X_1,\dots,X_n$ of $n$ correlated source symbols, and
using scheme 1, what is observed by an
eavesdropper is the coset valued sequence of random elements
$\{H(s_n(\bX))\}$, $H$ being the parity check
matrix. Since $\bX$ is a correlated source of symbols, there is
no a priori reason to expect that the coset valued process will
not be correlated. For example if $\bX$ forms a Markov chain,
then the coset valued process is a function of a Markov chain; although it will not, in general, form a Markov chain itself, it
will still have correlations. These correlations could effectively
reduce the list size that an eavesdropper must search and, consequently, reduce the effectiveness of the scheme. Reducing or eliminating
correlations in the coset valued process would counteract the
impact of this vulnerability.

Different approaches can be taken to resolve this issue. In general, the key to
reducing the effect of the correlation between codewords is to encode larger
block lengths. More precisely, let $\bX_1,\bX_2,\dots, \bX_N$ be $N$ blocks of
symbols produced by a Markov source, such that $\bX_i \in \Xn$ and
$p(\bX_1,\dots,\bX_N)= p(\bX_1)p(\bX_2|\bX_1)\dots p(\bX_N|\bX_{N-1})$. Instead
of  encoding each block individually, the transmitter can compute
$\bY^{nNR}=f(\bX_1,\dots,\bX_N)$.

The previous approach has the disadvantage of requiring long block lengths and
possibly high implementation complexity. We note, however, that the encoding
procedure over multiple blocks does not necessarily have to be performed
independently. For example, one possible approach for overcoming edge-effect
correlations between codewords is to define
$\bY_1=f(\bX_1,\bX_2),\bY_2=f(\bX_2,\bX_3),\dots$, and so forth. This approach
reduces the edge effects of correlation between codewords, in particular when
the individual sequences $\bX_i$ are already significantly long.

We note that, when probabilistic encryption \cite{katz_introduction_2007}  is required
over multiple blocks, the source encoded symbols in scheme 1 can be
combined with the output of a pseudorandom number generator (PRG) before being
multiplied by the parity check matrix. This would provide  the necessary randomization of the
output. The initial seed of the PRG can then be transmitted to the legitimate
receiver in phase II of scheme \ref{scheme:prac}.

\section{Applications and Practical Considerations}
\label{sec:practical}

The protocol outline presented in scheme \ref{scheme:prac}  is useful in
different practical scenarios, which are discussed in the following sections.
Most of the advantages of the suggested scheme stem from the fact that
list-source codes are key-independent, allowing content to be distributed  when
a key distribution infrastructure is not yet established, and providing an
additional level of security if keys are compromised before phase II in scheme
\ref{scheme:prac}.

\subsection{Content pre-caching}

As hinted earlier, list-source codes provide a secure mechanism for content
pre-caching when a key infrastructure has not yet been established. A large
fraction of the data can be  list-source coded and  securely transmitted before
the termination of the key distribution protocol. This is particularly
significant in large networks with hundreds of mobile nodes, where key
management protocols can require a significant amount of time to complete
\cite{eschenauer_key-management_2002}. Scheme \ref{scheme:prac} circumvents the
communication delays incurred by key compromise detection, revocation and
redistribution by allowing data to be efficiently distributed concurrently with
the key distribution protocol, while maintaining a level of security determined
by the underlying list-source code.

\subsection{Application to key distribution protocols}

List-source codes can also provide additional robustness to key compromise. If
the secret key is compromised before phase II of scheme \ref{scheme:prac}, the
data will still be as secure as the underlying list-source code. Even if a
(computationally unbounded) adversary has perfect knowledge of the key, until
the last part of the data is transmitted the best he can do is reduce the number
of possible inputs to an exponentially large list. In contrast, if a stream
cipher based on a pseudo-random number generator were used and the
initial seed was leaked to an adversary, all the data transmitted up to the
point where the compromise was detected would be vulnerable. The use of
list-source codes provide an additional, information-theoretic level of security
to the data up to the point where the last fraction of the message is
transmitted.  This also allows decisions as to which receivers will be allowed
to decrypt the data can be delayed until the very end of the transmission,
providing more time for detection of unauthorized receivers and allowing a
larger flexibility in key distribution.

In addition, if the level of security provided by the list-source code is
considered sufficient and the key is compromised before phase II, the key can be
redistributed \textit{without the need of retransmitting the entire data}. As
soon as the keys are reestablished, the transmitter simply encrypts the
remaining part of the data  in phase II with the new key.

\subsection{Additional layer of securi ty}

We also highlight that list-source codes can be used to provide an additional
layer of security to the underlying encryption scheme. The message can be
list-source coded  after encryption and transmitted in two phases, as in scheme
\ref{scheme:prac}. As argued in the previous point, this provides additional
robustness against key compromise, in particular when a compromised key can
reveal a large amount of information about an incomplete message (e.g. stream
ciphers). Consequently, list-source codes are a simple, practical way of
augmenting the security of current  encryption schemes.

One example application is to combine list-source codes with stream ciphers, as
noted in section V. The
source-coded message can be initially encrypted using a pseudorandom number generator
initialized with a randomly selected seed, and then list-source coded. The
initial random seed would be part of the encrypted message sent in the final
transmission phase. This setup has the advantage of augmenting the security of
the underlying stream cipher, and provides randomization to the list-source
coded message. In particular, if the LSC is based on MDS codes and assuming that
the distribution of the plaintext is nearly uniform, strong
information-theoretic symbol secrecy guarantees can be made about the
transmitted data, as discussed in section \ref{sec:security}.  Even if the
underlying PRG is compromised, the message would still be secure.

\subsection{Adjustable level of secrecy }

List-source codes provide a tunable level of secrecy, i.e. the amount of
security provided by the scheme can be adjusted according to the application of
interest. This can be done by appropriately selecting the size of the list ($L$)
of the underlying code, which determines the amount of uncertainty an adversary
will have regarding the input message. In the proposed implementation using
linear codes, this corresponds to choosing the size of the parity check matrix
$\mathbf{H}$, or, analogously, the parameters of the underlying error-correcting
code. In terms of scheme \ref{scheme:prac}, a larger (respectively smaller)
value of $L$ will lead to a smaller (larger) list-source coded message in phase
I and a larger (smaller) encryption burden in phase II.



\section{Conclusions}
\label{sec:conc}

In this paper we  introduced the concept of list-source codes, which are codes
that compress a source below its entropy rate. We derived fundamental bounds for
the rate list region, and provided code constructions that achieve these bounds.
List-source codes are a useful tool to understand how to perform encryption when
the (random) key length is smaller that the message entropy. In a nutshell, when
the key is small, we can reduce an adversary's uncertainty to a near-uniformly
distributed list of possible source sequences with an exponential (in terms of
the key length) number of elements by using list-source codes. We also
demonstrated how list-source codes can be implemented using standard linear
codes. 

Furthermore, a new information-theoretic metric of secrecy was presented, namely
$\epsilon$-symbol secrecy, which characterizes the amount of information leaked
about specific symbols of the source given an encoded version of the message. We
derived fundamental bounds for  $\epsilon$-symbol secrecy, and showed how these
bounds can be achieved using MDS codes when the source is uniformly distributed.
Finally, we discussed how list-source codes can be applied to practical
encryption schemes. 

\bibliographystyle{IEEEtran}
\bibliography{IEEEabrv,references}

\end{document}